# Improved Cryogenic Photodiode Optical Biasing for Low-Noise and Low-Jitter Superconducting Nanowire Single-Photon Detectors

**Jia-Hao Hu[1, 2, 3], Wei-Jun Zhang[1, 2, 3*], Wen-Shuo Yu[1, 2, 3], Yu-Ze Wang[1, 2, 3], Dong-Wei Chu[1, 2, 3], Ya-Tao Peng[4], Hui-Qin Yu[1, 2, 3], Pu-Sheng Yuan[1, 2, 3], Ling Wu[1, 2, 3], and Li-Xing You[1, 2, 3]**

[1]State Key Laboratory of Materials for Integrated Circuits, Shanghai Institute of Microsystem and Information Technology (SIMIT), Chinese Academy of Sciences (CAS), Shanghai 200050, China

[2]Shanghai Key Laboratory of Superconductor Integrated Circuit Technology, Shanghai Institute of Microsystem and Information Technology (SIMIT), Chinese Academy of Sciences, Shanghai 200050, China.

[3]Center of Materials Science and Optoelectronics Engineering, University of Chinese Academy of Sciences, Beijing 100049, China.

[4]State Key Laboratory of Analog and Mixed-Signal VLSI, Institute of Microelectronics, and FST ECE, University of Macau, Macau, China.

Contact author: [*]zhangweijun@mail.sim.ac.cn

**Abstract:** We experimentally demonstrate an improved optical biasing scheme for superconducting nanowire single-photon detectors (SNSPDs), which employs a cryogenic InGaAs–InP photodiode (PD) as a local bias source. It is found that, under illumination from a stable external light source, this PD generates a stable photocurrent in a cryogenic environment (~2.3 K), with fluctuations in the photocurrent primarily attributed to fluctuations in the incident optical power. Furthermore, by screening and effectively blocking stray photons leaking from the PD, which give rise to background dark counts, we have achieved an SNSPD exhibiting an ultra-low intrinsic dark count rate of $1 \times 10^{-4}$ cps. Utilizing this improved optical biasing technique, our SNSPD achieved performance comparable to that obtained under conventional electrical biasing: a system detection efficiency of 80.7%, a background dark count rate of 32.6 cps, and a minimum timing jitter of 57.5 ps. These results indicate that cryogenic-PD-based optical biasing serves as a viable, low-noise, and low-jitter alternative to traditional electrical biasing. Moreover, this work offers useful design guidance for the future development of PD-based low-noise bias sources and for the construction of all-photonic SNSPD systems tailored for high-precision quantum photonics applications.

## 1. Introduction

Single-photon detection with high efficiency, low noise, and low timing jitter is a key enabling technology for quantum information processing [1], optical quantum computing [2], deep-space optical communication [3], and photon-starved imaging [4]. Superconducting nanowire single-photon detectors (SNSPDs) have emerged as the state-of-the-art solution due to their near-unity system detection efficiency (SDE >98%), timing jitter at the picosecond level (e.g., tens of ps), and ultra-low dark count rates (DCR, in some cases $<10^{-3}$ cps) [5-7]. These characteristics make SNSPDs indispensable for applications requiring single-photon-level precision.

A stable and low-noise bias current source is crucial for achieving optimal performance of SNSPDs. The operation of SNSPDs relies on the formation of a resistive hotspot when an absorbed photon locally drives a section of the superconducting nanowire above its critical current density [8]. To maximize the SDE while minimizing the intrinsic DCR, the bias current ($I_b$) must be precisely maintained near, but below, the switching current ($I_{sw}$) typically at $I_b \approx 0.9-0.95 I_{sw}$. Small current fluctuations on the order of ~0.1μA would shift the detector's operating point, leading to excess dark counts, counting, or timing instability, thus degrading the operational stability and reliability.

Conventional SNSPD systems employ room-temperature constant-current or quasi-constant-current sources connected via long coaxial cables to the cryogenic stage [8]. However, such electrical biasing introduces non-negligible thermal load (typically tens to hundreds of μW per cable) [9], thermal noise coupling, and electromagnetic interference (EMI) through the bias cables. These effects become increasingly critical in large-scale SNSPD arrays [3] or on-chip photonic platforms [10], where wiring density and heat dissipation strongly limit scalability.

To address these limitations, photonic interconnects between room-temperature electronics and cryogenic detector platforms have been proposed [11-16]. Among them, optical (or photonic) biasing and readout schemes have recently emerged as promising alternatives [11, 12]. In such architectures, both bias control and signal transmission are implemented through optical links, eliminating the need for direct electrical wiring between the room-temperature control electronics and the cryogenic stage. This configuration effectively suppresses EMI, reduces thermal loading through wiring, and minimizes inter-channel crosstalk. Furthermore, optical interconnects are inherently compatible with integrated quantum photonic circuits, offering enhanced scalability, precise local control, and high-throughput signal distribution for large-scale superconducting device or detector systems.

In this context, cryogenic photodiodes (PDs) operating at temperatures below 4 K, such as InGaAs/InP heterojunction diodes, have been demonstrated as local optical-to-electrical converters that provide current bias for qubits[14, 17] or SNSPDs [11]. These PDs generate photocurrent proportional to the incident optical power, allowing a current source that directly drives the devices. For SNSPDs, system detection efficiencies (SDE ≈ 82.5%) comparable to electrical biasing have been shown by Thiele et al in 2022 [11].

Despite these advances, several critical issues remain unresolved. First, the reported detection noise—quantified by the DCR (~31kcps), and the timing jitter (~495 ps) remain significantly higher than those achieved with conventional electrical bias circuits [18]. Second, the fundamental question of whether a cryogenic PD can serve as a truly low-noise and stable current source for SNSPDs has not yet been fully clarified. A quantitative understanding of these interdependencies is essential for realizing fully photonic SNSPD systems.

In this work, we systematically investigate the performance of SNSPDs biased by both a cryogenic PD and a conventional low-noise electronic current source. We first characterize the photocurrent stability of the cryogenic PD. The intrinsic DCR of the optically biased SNSPD reveals that stray photon leakage from the PD can induce excess counts, while optimized light shielding can eliminate this effect. Then, key performance metrics of optical biasing SNSPD, including DCR, SDE, timing jitter, and maximal count-rate, are compared with electrical biasing under similar conditions.

## 2. Stability characterization of PD as a cryogenic current source

A cryogenic photodiode (MTPD1346D-010, Marktech Optoelectronics) housed in an FC connector flange, allowing direct fiber-to-device coupling, was mounted at the 2-K stage of the cryostat and employed as the bias current source for the SNSPD. The photocurrent generated by the PD follows the photoelectric relation [19]:

$$I_p = \eta \cdot q \cdot P_{opt}/(h \cdot v),$$

where $I_p$ is the photocurrent, $\eta$ the quantum efficiency of the PD, $q$ the electron charge, $P_{opt}$ the incident optical power, $h$ Planck's constant, and $v$ the optical frequency. Under constant $\eta$, $I_p$ exhibits a linear dependence on $P_{opt}$, implying that current variation directly related to the optical power fluctuations. Therefore, the photocurrent stability is expected to be mainly limited by the incident optical power.

To achieve stable and low-noise operation, a power-stabilized 1310-nm continuous-wave (CW) laser (model AQ2200-112, YOKOGAWA) was used as the optical excitation source. As shown in Fig. 1(a), the laser output was directed to the PD through a variable optical attenuator (81560A, Keysight) and a single-mode fiber, where the generated photocurrent was delivered to a room temperature picoammeter (Model 6485, Keithley) through a cryogenic coaxial cable and a room-temperature coaxial cable.

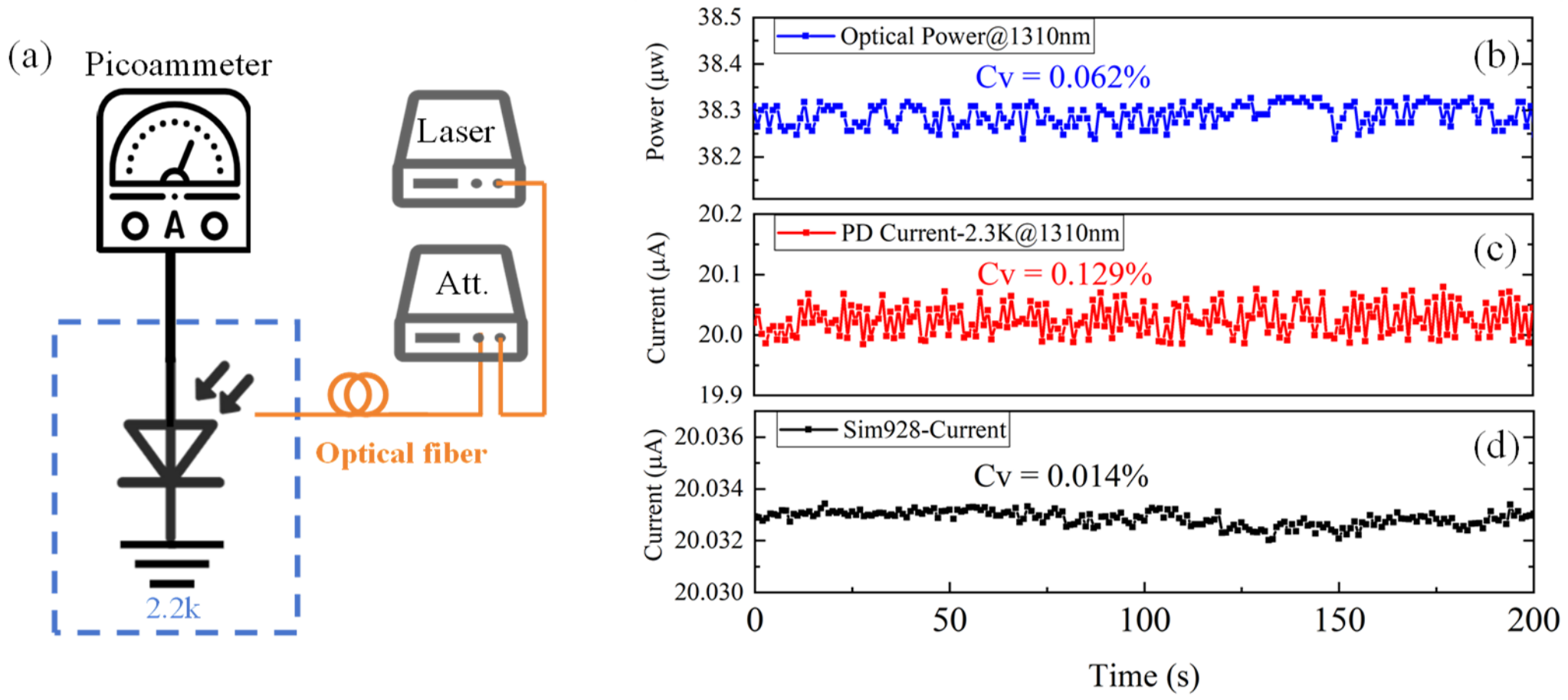

**Figure 1.** (a) Schematic of the experimental setup. Light from a stable room-temperature laser is delivered through a tunable attenuator (Att.) and an optical fiber to a photodiode (PD) mounted on the 2 K stage of the cryostat. The resulting photocurrent is measured using a picoammeter. (b) Optical power of the 1310 nm CW laser measured at 1 s intervals using a power meter. (c) Output photocurrent monitoring of the cryogenic PD, measured at 1 s intervals using the picoammeter. (d) Current stability of a SIM928 precision voltage source connected in series with a 10 kΩ resistor, measured at 1 s intervals using the same picoammeter.

As shown in Fig. 1b and 1c, the temporal fluctuations of the output optical power and the corresponding photocurrent were simultaneously monitored using a high precision optical power meter (81624B, Keysight) and the picoammeter, respectively. The dark current of the cryogenic PD under zero illumination (blocking by a shutter of the attenuator) was measured to be below 10 pA, confirming negligible thermal carrier generation at 2 K. For comparison, a conventional electrical bias circuit consisting of a SIM928 precision voltage source in series with a 10 kΩ resistor was characterized at room temperature using the same picoammeter [Fig. 1(c)].

To quantify the relative amplitude of current and power fluctuations, the coefficient of variation (Cv = $\sigma/\mu \times 100\%$), where, $\sigma$ denotes the standard deviation, and $\mu$ denotes the mean. The photocurrent fluctuations of approximately ±0.05 μA (Cv = 0.129%) corresponding to an incident optical power variation of ±0.044 μW (Cv = 0.062%). The photocurrent Cv was roughly twice that of the optical power, primarily attributed to power instability introduced by the FC/PC fiber connector interface of the input fiber. In contrast, the electrical bias circuit demonstrated lower fluctuations of ±0.008 μA (Cv = 0.014%), representing about one order of magnitude higher current stability.

These results shows that the dominant fluctuations in the photocurrent originates from residual optical power fluctuations of the laser and instability along the optical path. Further suppression could be achieved by employing an ultra-stable optical source, fusion-spliced fiber coupling, and thorough connector cleaning to minimize interface-induced mechanical perturbations and coupling instability.

## 3. Intrinsic DCR under Cryogenic Optical Bias

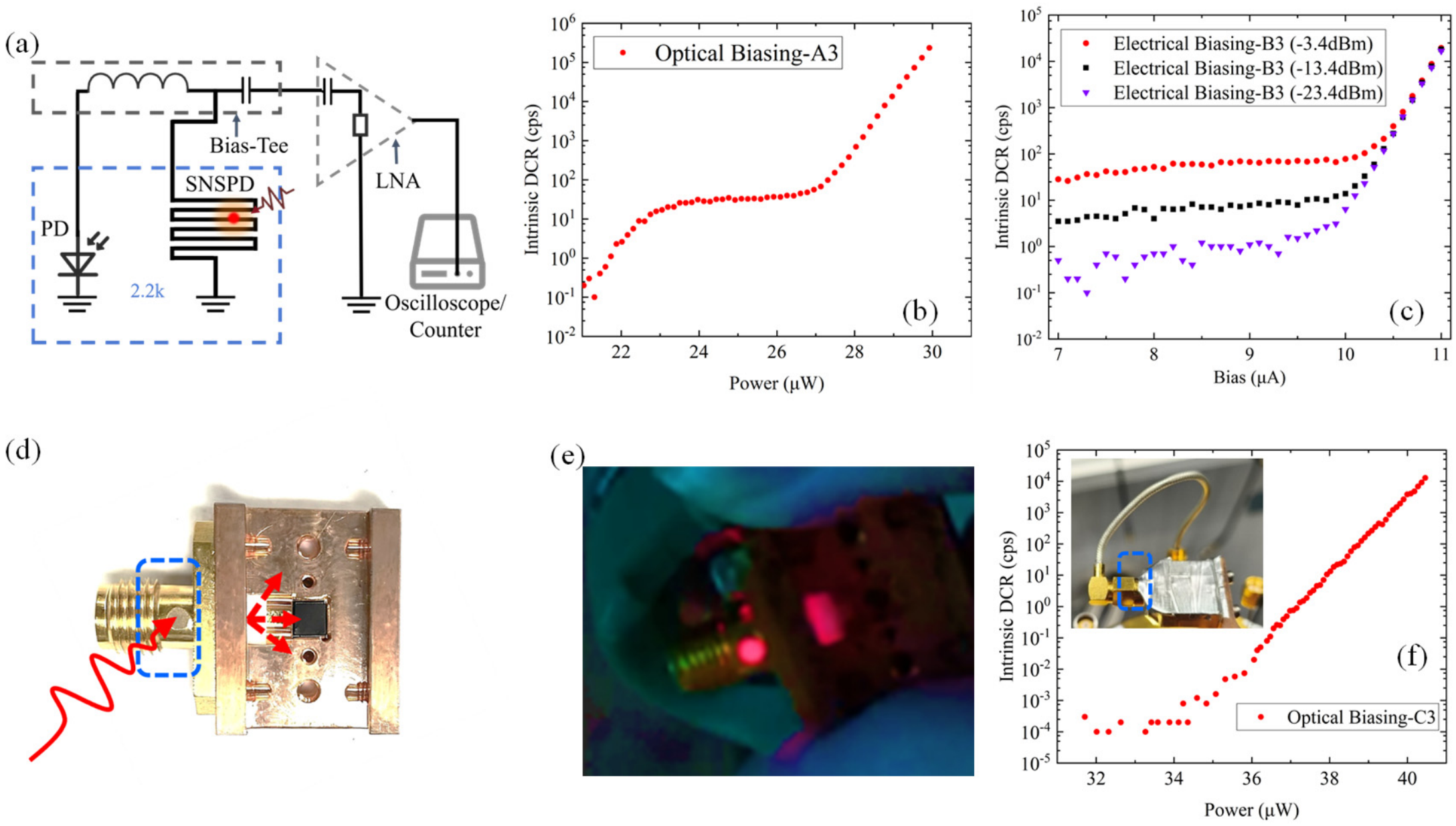

**Figure 2.** (a) Schematic of the optically biased SNSPD and its performance characterization setup. (b) Intrinsic dark count rate (DCR) as a function of bias optical power for a conventional radiation-shielded SNSPD (Device B3) under cryogenic PD optical biasing. (c) Intrinsic DCR as a function of bias current for a conventional radiation-shielded SNSPD (Device A3) under electrical biasing using a SIM928 source, while the cryogenic PD on the same stage is illuminated with different optical powers. (d) Illustration of a leakage path in the package block (indicated by the blue dashed box), where the white dielectric insulator of the SMA connector allows stray photons to reach the SNSPD chip. (e) Visualization of photon scattering inside the package using a red laser pointer, with illumination directed at the dielectric insulator of the SMA connector. (f) Intrinsic DCR of the fully radiation-shielded SNSPD (Device C3) under PD optical biasing. Inset: Photograph of the fully shielded SNSPD packaging, where the detector is enclosed in a copper housing. The dielectric insulator and the four sidewalls of the package are wrapped with aluminum foil tape.

The intrinsic DCR of an SNSPD represents photon-independent counts arising from internal physical processes, such as thermal fluctuations, current-assisted vortex motion. These processes can be further enhanced by local geometric constrictions in the nanowire [20], and therefore reflects the detector's fundamental noise floor. Experimentally, the intrinsic DCR is measured with the SNSPD enclosed in a light-tight package wrapped with aluminum foil and operated without any optical fiber connection, thereby eliminating contributions from external photons [21].

The intrinsic DCR of the SNSPD was measured under optical biasing provided by a cryogenic PD. Specifically, the photocurrent generated by the PD at the 2 K stage was routed through a cryogenic coaxial cable to the DC port of a room-temperature bias tee (ZX85-12G-S, Mini-Circuits) and then returned to the detector through a second coaxial cable, thereby serving as the bias current for the SNSPD. The RF port of the bias tee directed the detector output to a low-noise amplifier (LNA650, RF Bay Inc.), followed by either a photon counter (SR400, Stanford Research Systems) or an oscilloscope (MOSV204A, Keysight) for signal analysis. The experimental configuration is shown in Fig. 2(a). This arrangement allowed the same room-temperature bias and readout circuitry to be used for both optical and conventional electrical biasing measurements, thereby minimizing systematic differences between the two biasing schemes.

During the initial measurements, the SNSPD package was wrapped with aluminum foil tape following a standard radiation-shielding procedure, while the cryogenic PD was connected via a black-jacketed single-mode fiber. Devices B3 and A3 were fabricated from the same wafer and exhibited similar nanowire geometries and switching currents. Therefore, the two nominally identical devices were used to enable a controlled comparison between optical and electrical biasing.

However, unlike the typical log-linear dependence of the intrinsic DCR on bias current [21], an anomalous DCR floor ranging from approximately 0.1 to 100 cps was observed in the low-bias region under optical biasing [Fig. 2(b), Device A3]. To investigate the origin of this anomaly, Device B3 was biased electrically using a conventional current source (SIM928), while the cryogenic PD mounted on the same temperature stage was illuminated with different optical powers [Fig. 2(c)]. The laser output power before attenuation was approximately 3.6 dBm, and optical attenuations of 7, 17, and 27 dB were applied. The stage temperature remained essentially unchanged throughout the measurements, ruling out thermal effects as the source of the excess counts. As the optical power incident on the PD was reduced, the DCR floor decreased monotonically. This behavior indicates that stray photons originating from the PD, rather than temperature variations, were responsible for the excess dark counts.

Further inspection revealed that the leakage path originated from the white plastic dielectric insulator of the SMA connector [Fig. 2(d)]. When illuminated with a visible red laser [Fig. 2(e)], light was observed to propagate and scatter within the dielectric, providing an optical path into the detector enclosure. To suppress this effect, the connector and all residual gaps in the copper housing were carefully sealed using aluminum foil tape. After this modification, the DCR recovered its typical log-linear dependence on bias current and became independent of the PD illumination power [Fig. 2(f)].

With the optimized radiation-shielded configuration, the optically biased SNSPD (Device C3) exhibited an ultra-low intrinsic DCR of approximately $1\times10^{-4}$ cps, comparable to that achieved under conventional electrical biasing. These results demonstrate that a cryogenic PD can serve as a stable and low-noise current source for SNSPD operation. Additional shielding strategies, such as thickening the black-jacketed fiber and enclosing the PD package in a metallic housing, may further suppress stray-photon leakage.

It is worth noting that light leakage through the dielectric insulator of SMA connectors has received little attention in previous SNSPD studies. A likely reason is that, in conventional SNSPD systems, the ambient blackbody photon flux inside the cryostat is relatively low, making the probability of photons propagating through the dielectric and reaching the detector negligible. In the present work, however, the fiber-coupled cryogenic PD introduces a substantially larger local photon flux. Consequently, photons escaping from the coupling structure can propagate through the cryostat and eventually reach the SNSPD, rendering this leakage path experimentally observable through an increase in the measured DCR.

We further observed that a noticeable excess background DCR persisted even when the fiber-coupled PD was positioned more than 10 cm away from the SNSPD package under conventional shielding conditions. This suggests that, within the enclosed cryostat equipped with gold-plated radiation shields, multiple scattering of stray photons can still result in a finite probability of coupling into the SNSPD through the dielectric insulator of the SMA connector. The present work demonstrates that optical leakage paths that are negligible in conventional SNSPD systems can become significant when local optical sources are introduced into the cryogenic environment. This observation further highlights the extraordinary sensitivity of SNSPDs to weak environmental optical noise and provides useful design guidance for future cryogenic photonic systems employing optical power delivery or optical control.

## 4. Detector Performance characterization of Optically Biased SNSPDs

To establish a correspondence between the optical-bias and electrical-bias current axes, the optical power incident on the cryogenic photodiode was first calibrated using a power meter and subsequently adjusted with a variable optical attenuator. The resulting photocurrent was measured using a picoammeter. By correlating the incident optical power with the corresponding photocurrent, the optical-bias operating points were converted into equivalent bias currents, enabling a direct comparison between optical and electrical biasing performance.

Figure 3 shows the photocurrent response of the photodiode measured at 2.2 K, which was used to establish this calibration relationship. The measured responsivity of approximately 0.349 A/W is lower than typical reported values (~0.6–0.7 A/W at 1550 nm) [8], which is likely due to additional coupling losses in the cryogenic optical packaging, including Fresnel reflections at the fiber-to-detector interface and residual optical misalignment.

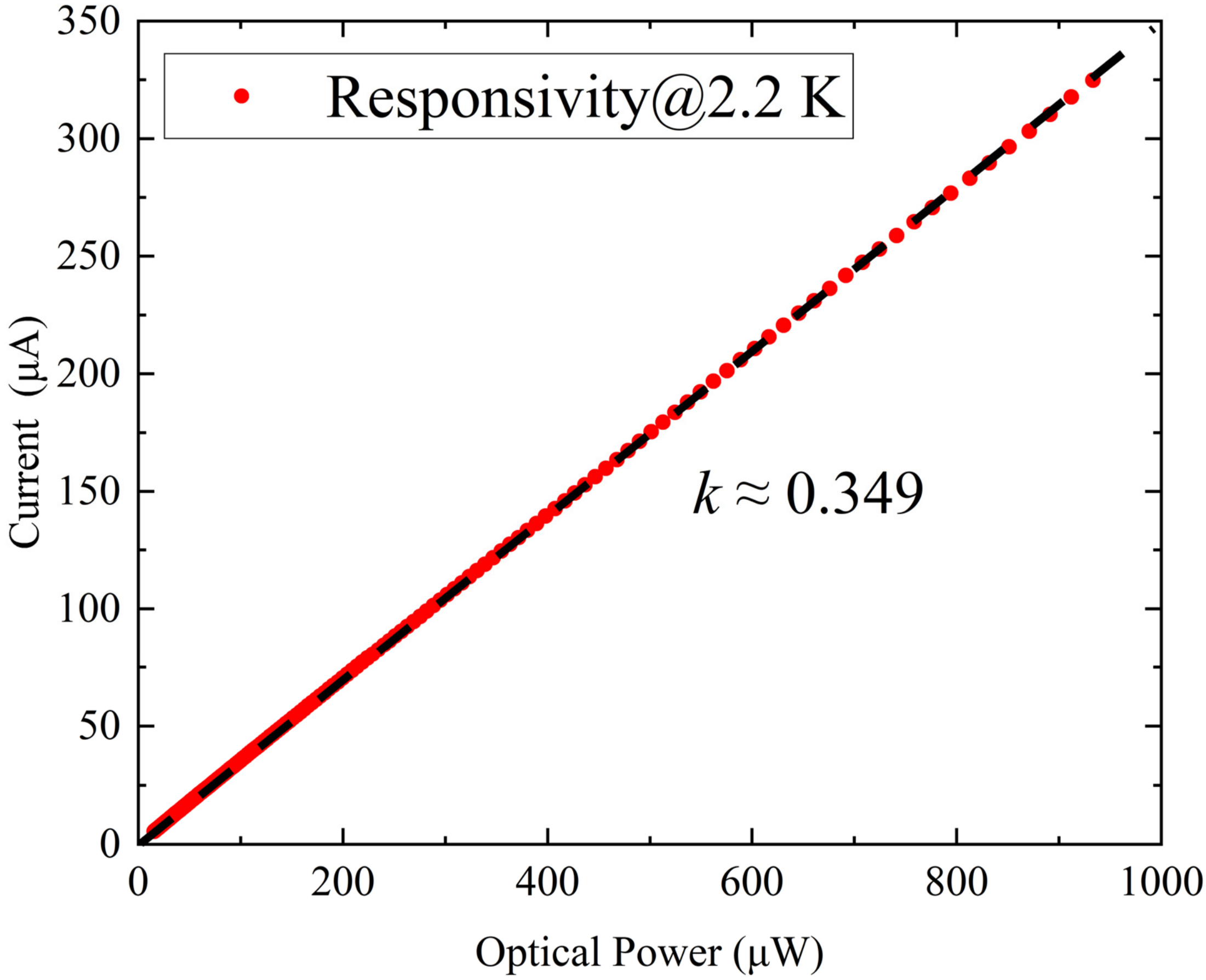

**Figure 3.** Photocurrent response of the photodiode at 2.2 K as a function of incident optical power (red dots), along with a linear fit (black dashed line). The linear fit yields a coefficient of determination $R^2 \approx 0.9999$ and a responsivity of ~0.349 A/W.

Next, the detection performance of the SNSPD was characterized. An attenuated 1550 nm pulsed laser (FPL-01CFFSIT, Calmar Laser) was used as the single-photon excitation source. The optical power was adjusted to provide a photon flux of approximately $1.0\times10^5$ photons $s^{-1}$. The SDE was calculated using SDE = (PCR − DCR)/$N_p$, where PCR is the photon count rate, and $N_p$ is the incident photon flux. The estimated uncertainty in the SDE measurement was approximately 3%. The timing jitter was measured using the high-speed oscilloscope, with the synchronization signal from the pulsed laser used as the reference trigger [22]. The full width at half maximum (FWHM) of the timing distribution was extracted as the jitter value.

As shown in Fig. 4(a–d), optical and electrical biasing measurements were performed sequentially on the same SNSPD (Device D3) to enable a direct performance comparison. Under optical biasing, Device D3 exhibited an SDE of 80.7%, a DCR of 32.6 cps, and a timing jitter of 57.5 ps. These values are comparable to those obtained under conventional electrical biasing, which yielded an SDE of 81.0%, a DCR of 31.8 cps, and a timing jitter of 57.0 ps. The PCR curves as a function of input optical power nearly overlap for the two biasing schemes, indicating that the SNSPD operates stably under optical biasing without premature latching. These results demonstrate the feasibility of cryogenic PD-based optical biasing for high-performance SNSPD operation.

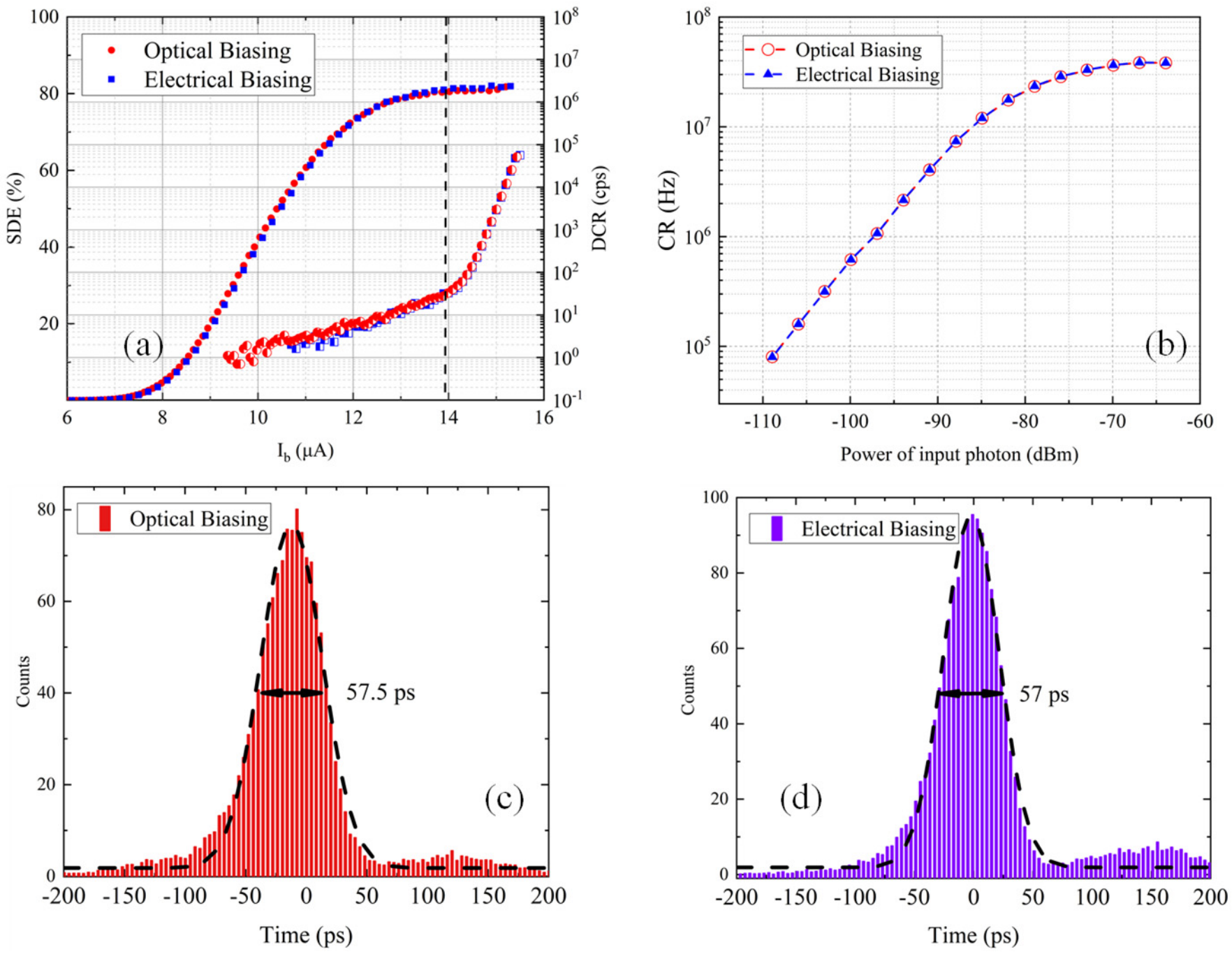

**Figure 4.** (a) System detection efficiency (SDE) and DCR of the SNSPD under optical (red) and electrical (black) biasing. (b) Photon count rate (PCR) of the SNSPD under optical and electrical biasing. (c) Timing jitter of the SNSPD under electrical biasing. (d) Timing jitter (FWHM of the timing distribution) of the SNSPD under optical biasing.

To further assess bias stability, we compared SDE fluctuations over 900 s at two key bias current points: the saturation plateau and the maximum-slope of the current dependent SDE curve (Fig. 5a–e). In the saturation regime ($0.93I_{sw}$), both optical and electrical biasing showed equivalent stability (Cv≈ 0.37%). Near the transition point ($0.60I_{sw}$), optical bias exhibited slightly larger fluctuation (Cv ≈ 0.66%) but remained within the acceptable range for stable operation.

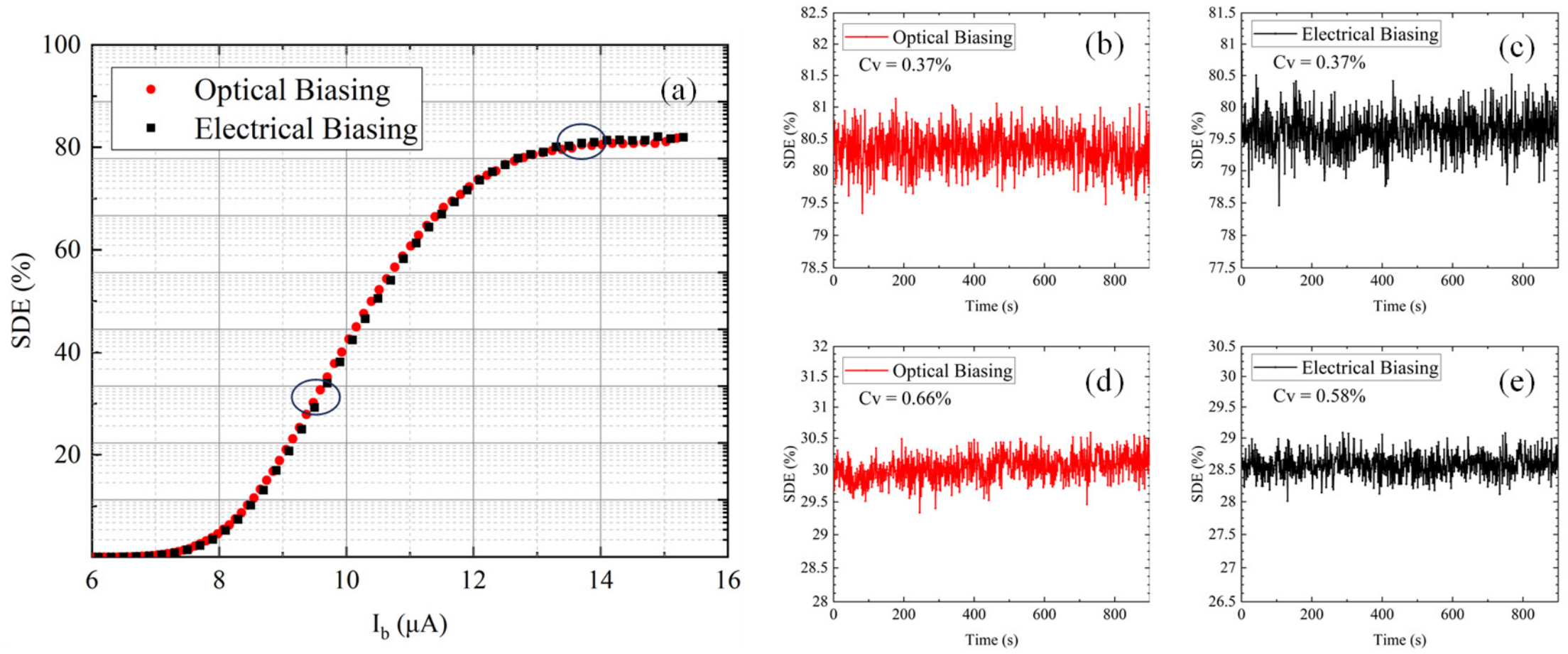

**Figure 5.** Comparison of SDE stability under optical and electrical biasing. (a) SDE vs bias current curves under optical (red dots) and electrical (black squares) biasing. The stability of SDE and bias current was evaluated at two characteristic operating points of the SDE curve: the saturation plateau (b, c) and the maximum-slope region (d, e). (b) and (d) correspond to the saturation plateau, while (c) and (e) correspond to the maximum first-order derivative (steepest slope) region.

## 5. Conclusion

We demonstrated and analyzed an improved optical biasing scheme for SNSPDs using a cryogenic PD as a local current source. Operating at 2.3 K, the PD provides a stable bias current with a coefficient of variation of 0.37%, primarily governed by fluctuations in the incident optical power, arising from the light source and optical delivery path. After implementing comprehensive optical shielding, we achieved an intrinsic DCR as low as $1 \times 10^{-4}$ cps for an optically biased SNSPD.

Furthermore, the optically biased SNSPD exhibited a system detection efficiency of 80.7%, a background DCR of 32.6 cps, and a timing jitter of 57.5 ps, all comparable to those obtained under conventional electrical biasing. These results demonstrate that cryogenic PD-based optical biasing can serve as a low-noise and low-jitter biasing approach for high-performance SNSPD systems, enabling a pathway toward fully photonic, cryogenically integrated quantum detection platforms.

The overall system timing jitter could be further reduced by employing a cryogenic amplifier. Beyond single-device demonstrations, the integration of cryogenic PDs as distributed optical bias sources offers a scalable route toward SNSPD arrays and photonic–superconducting hybrid architectures.

**Acknowledgments:** This work is supported by the Quantum Science and Technology-National Science and Technology Major Project (Grant No. 2023ZD0300100), the National Natural Science Foundation of China (Grant Nos. 62371443 and 6230400). W.-J. Zhang is supported by an outstanding member of the Youth Innovation Promotion Association, CAS (Y2023071). The authors thank Xiao-Yu Liu for technical support electron beam lithography. An AI assistant (DeepSeek V3.2-Exp) was used for English language refinement, and all outputs were subsequently reviewed and verified by human editors.

**Author contributions:** W.-J.Z. and J.H.H. conceived and designed the experiments, W.-J.Z. and J.H.H. fabricated the samples. J.H.H. conducted the experiments and collected the data. J.H.H., Y.T.P., and W.-J.Z. analyzed the data and prepared the manuscript. All authors discussed the results and reviewed the manuscript.

**Competing interests:** The authors declare no competing interests.

**Data and materials availability:** All data that support the findings of this study are included within the article (and any supplementary files).